\newcommand{\fref}[1]{Fig.~\ref{#1}}

\newcommand{\ket}[1]{\ensuremath{|#1\rangle}}

\newcommand{\TLmatrix}[2]{\begin{array}{ccc} 0 & #1 & 0 \\ #1 & 0 &  #2 \\ 0 &  #2 & 0 \end{array}}
\documentclass[epj]{svjour}
\usepackage{graphics}
\usepackage{color}
\usepackage{soul}

\begin{document}
\title{Coherent injecting, extracting, and velocity filtering of neutral atoms in a ring trap via spatial adiabatic passage}
\titlerunning{Coherent injecting, extracting, and velocity filtering of neutral atoms in a ring trap}
\author{Yu. Loiko\inst{1} \and V. Ahufinger\inst{1} \and R. Menchon-Enrich\inst{1} \and G. Birkl \inst{2} \and J. Mompart\inst{1}
}                     
%
%
\institute{Departament de F\'{\i}sica, Universitat Aut\`{o}noma de Barcelona, E-08193 Bellaterra, Spain 
\and Institut f\"ur Angewandte Physik, Technische Universit\"at Darmstadt, Schlossgartenstra\ss e 7, 64289 Darmstadt, Germany}
\date{Received: \today / Revised version: date}
%
\abstract{
We introduce here a coherent technique to inject, extract, and velocity filter neutral atoms in a ring trap coupled via tunneling to two additional waveguides. By adiabatically
following the transverse spatial dark state, the proposed technique allows for an efficient and robust velocity dependent atomic population transfer between the ring and the input/output waveguides.
We have derived explicit conditions for the spatial adiabatic passage that depend on the atomic velocity along the input waveguide as well as on the initial population distribution among the transverse vibrational states. The validity of our proposal has been checked by numerical integration of the corresponding 2D Schr\"odinger equation with state-of-the-art parameter values for $^{87}$Rb atoms and an optical dipole ring trap.  
} 
\maketitle
\section{Introduction}
\label{intro}

Recent advances in the preparation and manipulation of ultra-cold neutral atoms have enabled a large number of high precision applications in different disciplines such as atom optics, quantum metrology, quantum computation, and quantum simulation \cite{general}. Neutral atoms can be trapped in a large variety of potential geometries, among which ring traps present some unique features, such as possessing periodic boundary conditions, that make them ideal candidates to investigate quantum phase transitions \cite{QPT}, Sagnac interferometry \cite{Sagnac}, stability of persistent currents and superconducting quantum interference devices \cite{cur_gen}, propagation of matter wave solitons and vortices \cite{sol_gen}, cold collisions \cite{ringmis}, artificial electromagnetism \cite{art}, etc. Bright optical (attractive) ring traps for neutral atoms have been experimentally reported with far-detuned optical dipole beams propagating through annular microlenses \cite{Birkl1}, with a combination of TOP traps and Gaussian beams \cite{Ryu07}, with Laguerre--Gauss (LG) modes \cite{Wri00,Fra07,Ram12}, and by means of spatial and acousto-optic modulators \cite{Sch08}, while dark (repulsive) optical rings have been realized by an appropriate superposition of Laguerre--Gauss laser modes \cite{Fra07}. In addition, time-averaged adiabatic potentials \cite{Les07,She11} are a very interesting alternative to generate versatile and smooth trapping potentials, including ring geometries.

In this paper, we discuss a coherent method for injecting neutral atoms into, extracting them from, and velocity filtering them in a ring trap, based on the spatial adiabatic passage technique (SAP). The SAP technique is the matter wave analog \cite{Eck04} of the stimulated Raman adiabatic passage (STIRAP) technique \cite{STIRAP} and has been previously proposed for efficiently transporting single atoms \cite{Eck04,atoms} and Bose--Einstein condensates \cite{BECs} between the two extreme traps of a triple-well potential, as well as for quantum tomography \cite{Loi11}. Recently, SAP for light propagating in a system of three coupled optical waveguides \cite{SAPlight} and a SAP-based spectral filtering device \cite{Men2} have been experimentally reported. 

\section{Physical system}

\begin{figure}[t]
\resizebox{0.45\textwidth}{!}{
\includegraphics{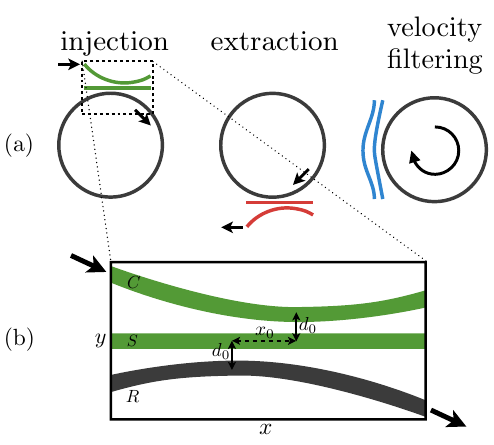}
}
\caption{
(Color online) 
(a) Schematic representation of the physical system consisting of a ring trap and two dipole waveguides for injecting neutral atoms into, extracting them from, and velocity filtering them in the ring waveguide. 
(b) Potential geometry for the injection protocol where the two dipole waveguides and part of the ring are modeled as three coupled waveguides. The input (curved) waveguide and the ring waveguide are denoted by $C$ and $R$, respectively, and correspond to two circularly bent waveguides, while the central straight waveguide is denoted by $S$. The distance $d_0$ accounts for the minimum $y$ separation between adjacent waveguides while $x_0$ gives the $x$ distance between the two positions of mimimum separation.}
\label{fig1}
\end{figure}

The physical system under consideration consists of a ring trap and two optical dipole waveguides coupled via tunneling in the three configurations shown in \fref{fig1}(a). In each case, the two waveguides coupled to the ring can be switched on or off at will by simply turning on or off the laser field that generates them. As a consequence, injection, extraction and velocity filtering of neutral atoms can be applied selectively when needed. The technique consists of adiabatically following a particular energy eigenstate of the system, the so-called transverse spatial dark state. Depending on the atomic velocity, an efficient and robust transfer of atoms between the ring and the outermost waveguide or vice versa takes place, with its performance surpassing the case of simply spatially overlapping the ring and the input/output waveguides. 

As a first configuration, we study the injection of a single cold neutral atom of mass $m$ into the ring trap by modeling the geometry of the physical system as a two-dimensional (2D) optical potential formed by three coupled dipole waveguides, see Fig.~1(b), with the ring and input traps being described as a segmented circular waveguide. The dynamics is governed by the 2D Schr\"odinger equation:
\begin{equation}
i \hbar\frac{\partial }{\partial t}\psi (x,y)=\left[
- \frac{\hbar^2}{2m}
\nabla^2 
+ {V}\left( x,y \right)\right ] \psi (x,y), \label{Eq_Schrodinger_2D}
\end{equation}
where $\nabla^2$ is the 2D Laplace operator and $V(x,y)$ is the waveguides trapping potential.
The transverse confinement for each waveguide is modeled by a harmonic potential of identical ground state width $\alpha \equiv \sqrt{\hbar /(m\omega _{\bot})}$, with $\omega _{\bot}$ being the transverse trapping frequency. With respect to the longitudinal width of the single-atom matter wave, we assume that the ring perimeter is much larger than the longitudinal width of the atomic wave packet and, therefore, azimuthal resonant frequencies of the ring do not play any role in the dynamics.

The geometry of each waveguide depends parametrically on $x$ through the variation of the corresponding waveguide center position $y_i=y_i (x)$ with $i=C,S,R$ labeling the curved, straight, and ring waveguides, respectively (Fig.~1(b)). 
Thus, the triple waveguide potential can be written as truncated harmonic potentials:
\begin{equation}
V\left( x,y\right) = \hbar \omega_{\bot}
\min_{i=C,S,R} \left[
\left( y-y_{i}\left( x\right) \right) ^{2}
/ 2 \alpha_{i} ^{2} \left( x \right) \right], 
\label{waveguide_potential}
\end{equation}
where $ \alpha_{i}^{2} \left( x \right) / \alpha ^{2} 
\approx1+\left| \partial y_{i}\left( x\right) / \partial x \right|^{2}$, with $\alpha_i$ being the width of the vibrational ground state of the $i$-th waveguide. Since the $y$-separation between the waveguides slowly varies along the $x$-axis, the motion of an atom with longitudinal velocity $v_{x}$ along any of the waveguides 
can be decoupled from the vibrational motion in the transverse direction.
Thus, the system is effectively reduced in the $y$-direction to a one-dimensional (1D) triple well potential
with the tunneling rates depending on the separation between the waveguides that, in the atom's own reference frame, are varied in time according to $y_i \left( x = v_x t \right)$. 

Note that truncated harmonic potentials are usually considered in the literature \cite{Eck04} since they allow for the analytical derivation of the tunneling rates. Although Gaussian or P\"oschl--Teller potentials describe much more accurately the experimental trapping potentials, it has been shown \cite{Loi11} that the adiabatic passage of a single atom between the outermost traps of a triple-well potential leads to qualitatively similar results for truncated harmonic and P\"oschl--Teller potentials. This is due to the fact that SAP does not need for an accurate control on the parameter values, i.e., even significant variations in the potential profiles do not qualitatively change the main physics of the process, as long as the initial and final matter wave vibrational states are resonant and the adiabaticity condition is fulfilled.

\section{Velocity filtering mechanism}

Assuming that the energy separation between the different vibrational states of each transverse harmonic potential 
is large enough to avoid crossed tunneling among different vibrational states and that there is no significant coupling between the two outermost waveguides, the transverse Hamiltonian $H_{\bot}$ of the system can be approximated to $H_{\bot}=H_0\oplus H_1 \oplus \dots \oplus H_n \oplus \dots$ \cite{Loi11} with:
\begin{equation}
H_{n} = \hbar \left( \TLmatrix{{ J}^{CS}_{n}(x)}{{ J}^{SR}_{n}(x)} \right),
\label{eqn:Ham_n}
\end{equation}
where $ J^{ij}_{n}$ is the tunneling rate between the $n$ vibrational states of two adjacent waveguides $i$ and $j$. 

After diagonalizing each Hamiltonian $H_n$, a set of three transverse energy eigenstates for each $n$ is obtained. One of them involves only the vibrational states of the two extreme waveguides and is known as the transverse spatial dark state \cite{Eck04}:
$ \ket{D_n ({\theta_n})}=\cos\theta_n \ket{n}_C - \sin\theta_n \ket{n}_R $, 
where $\theta_n$ is defined as $\tan \theta_n \equiv  J_n^{CS} /  J_n^{SR}$.
For a given $n$, SAP of a neutral atom between the two most separated waveguides 
can be achieved by adiabatically following the spatial dark state. This transfer process implies a smooth spatial variation of the tunneling rates such that
$\theta_n$ slowly changes from 0 to $\pi/2$.
In the protocol for injecting the atom from the left arm of the input waveguide $C$ to the ring $R$, the two initially empty waveguides, $S$ and $R$,
are first approached and separated along $x$ and, with an appropriate spatial delay $x_0$, the 
$C$ and $S$ waveguides are approached and separated (Fig.~1(b)).
This coupling sequence is performed to adiabatically follow the transverse spatial dark state $\ket{D_n ({\theta_n})}$ and is known as the counterintuitive coupling sentence \cite{STIRAP}. In fact, the spatial delay $x_0$ will be the key parameter in the geometry to guarantee the adiabaticity in the propagation dynamics of the atomic matter wave along the three coupled waveguides system. Note, however, that if the atom is injected from the right arm of the input waveguide C it will not follow a counterintuitive sequence of the couplings and it will be in a combination of transverse energy eigenstates which, in fact, will result in the spreading of the atomic probability density between the waveguides.

\begin{figure}[t]
\resizebox{0.4\textwidth}{!}{
\includegraphics{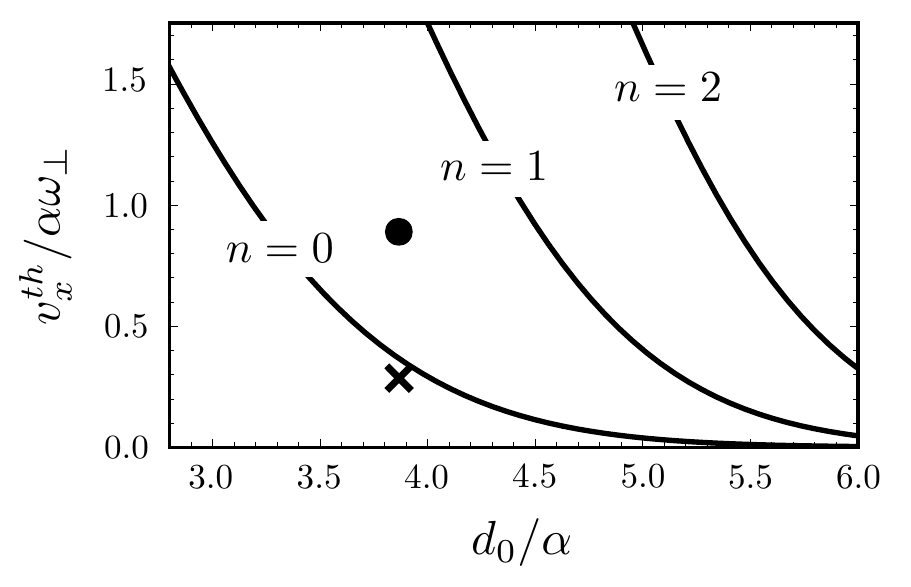}
}
\caption{
Longitudinal threshold velocity $v_x^{th}$ given by Eq.~(4) for the injection protocol as a function of the minimum waveguides separation $d_0$ for different transverse vibrational states $n$. In all cases $x_0=50\alpha$ and $A=10$. The cross and the circle mark the parameters used in the numerical simulations of Figs.~3(a) and 3(b), respectively.}
\label{fig2}
\end{figure}

Assuming identical transverse harmonic potentials, for a given $n$, the maximum tunneling rate between the curved and the straight, and between the straight and the ring waveguides will be equal, 
$ J^{CS}_n (- x_0/2)= J^{SR}_n (x_0/2)$ $(\equiv  J_n^{{\rm max}})$, corresponding to a separation $d_0$ between the two closest waveguides. 
The `global' adiabaticity condition for SAP reads $\sqrt{2} J_n^{{\rm max}}T > A$, 
where $T=x_0/v_x$ and $A$ is a dimensionless constant that takes a value around 10 for optimal parameter values \cite{STIRAP}, i.e., for a spatial delay between the couplings around $\sqrt{2}$ times the Gaussian-like widths of the couplings. Therefore, SAP will succeed for longitudinal atomic velocities fulfilling:
\begin{equation}
v_x < v_x^{th} \equiv \sqrt{2} J_n^{{\rm max}}x_0 /A .
\end{equation} 
Hence, by appropriately engineering the dimension $x_0$ of the interaction region and the tunneling rates through the minimum separation $d_0$, it is possible to implement a velocity filter such that atoms with a longitudinal velocity below the threshold velocity given by Eq.~(4) will be adiabatically transferred into the ring. As the tunneling rates depend on the transverse vibrational state $n$, the proposed adiabatic transfer technique can also be used for filtering vibrational states.
       
We have applied the recurrence Gram--Schmidt orthogonalization \cite{Loi11,Eck02} and Holstein--Herring methods \cite{HH} to obtain analytical expressions for the tunneling rates $ J_n^{{\rm max}}$. From these tunneling rates, we have plotted in Fig.~2 the longitudinal threshold velocity $v_x^{th}$ as a function of the minimum distance $d_0$ with $x_0=50\alpha$ and $A=10$. Clearly, since the threshold velocity increases when the tunneling rate increases, higher values for the threshold velocity are achieved at short distances $d_0$. It is also shown in this figure how the threshold velocity increases for increasing values of $n$, as expected since the tunneling rate grows with $n$.
Therefore, for a given distance $d_0$ between the waveguides, higher transverse vibrational states have higher coupling rates. Thus, it is possible to select particular parameter values for the geometry of the three coupled-waveguides system, e.g., the spatial delay $x_0$ between the waveguides, such that higher vibrational states fulfill Eq.~(4) and perform spatial adiabatic passage while lower vibrational states are not transferred due to their weaker coupling rates.

\begin{figure}[t]
\resizebox{0.5\textwidth}{!}{
\includegraphics{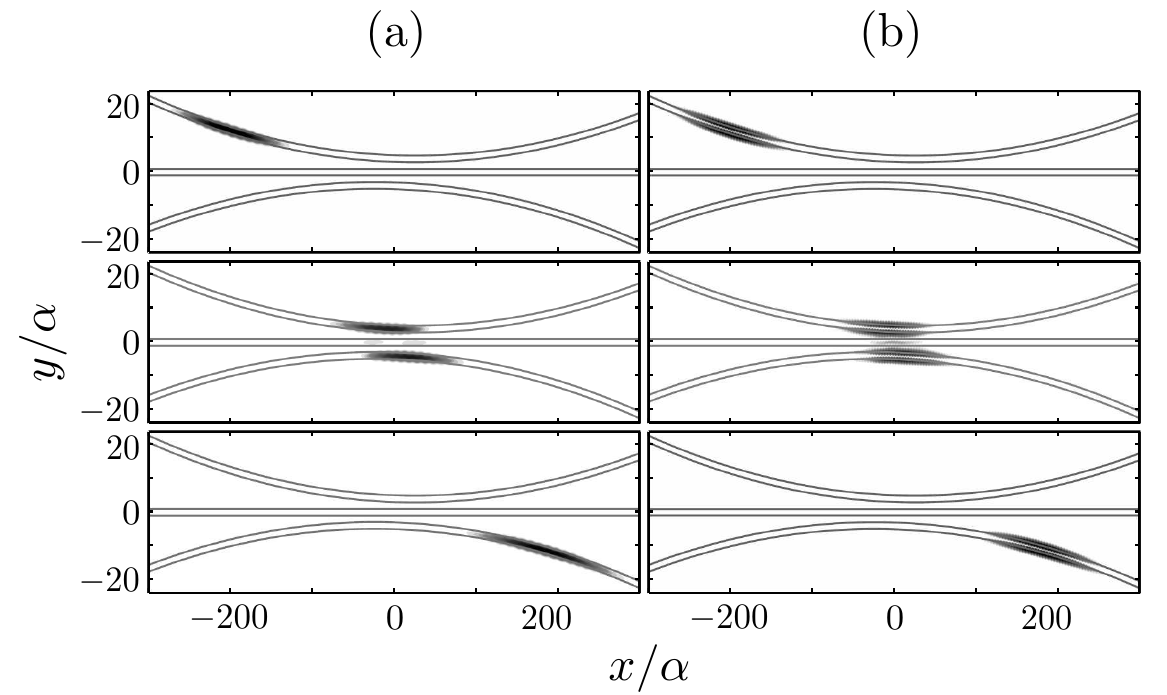}
}
\caption{
2D numerical simulations showing the atomic probability distribution at three different consecutive times for an input matter wave packet with (a) an initial velocity $v_x=0.3\alpha\omega_\bot $ at $n=0$, and (b) an initial velocity $v_x=0.9\alpha\omega_\bot $ at $n=1$. In both cases $d_0=3.9\alpha$, $x_0=50\alpha$, with the radius of the ring and curved waveguides being $r=3000\alpha$.}
\label{fig3}
\end{figure}

\section{Numerical results}

\subsection{Injection protocol}

Up to here, we have described the main ideas of the proposal by simplifying the initial 2D problem as a transverse 1D triple well potential with position dependent tunneling rates and analyzing the problem in terms of the transverse Hamiltonian. In order to be more realistic, we now investigate the injection protocol by a direct numerical integration of the 2D Schr\"odinger Eq.~(1) for $^{87}$Rb atoms with the geometry shown in Fig.~1(b) and the transverse potential given in Eq.~(2). Figs.~3(a) and 3(b) show three consecutive snapshots of the atomic probability distribution for the $n=0$ and $n=1$ transverse vibrational states, respectively, for the configurations marked in Fig.~2 by a cross (for Fig.~3(a)) and a dot (for Fig.~3(b)). They correspond to input velocities below the respective threshold velocities $v_x^{th}$, indicating that the global adiabaticity conditions for SAP are fulfilled.  
Thus, in both cases the atom is injected into the ring with a very high probability. Fig.~4 shows the numerically calculated region (in white) for which the transfer probability into the ring is higher than $97\%$ for the ground, Fig.~4(a), and first excited, Fig.~4(b), transverse vibrational states as a function of the input velocity and the minimum distance between waveguides, in good agreement with the threshold velocity expression derived in Eq.~(4) (solid lines). In addition, we have numerically confirmed that, for appropriate parameter sets, it is possible to perform filtering of the transverse vibrational states using the fact that the tunneling rates depend on $n$.

\begin{figure}[t]
\resizebox{0.5\textwidth}{!}{
\includegraphics{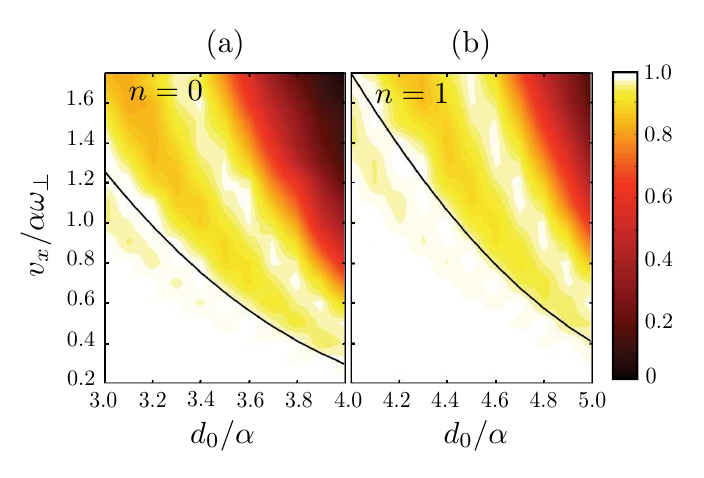}
}
\caption{
Numerically calculated probability for population injection into the ring waveguide as a function of the input velocity $v_x$ and the minimum separation $d_0$ for an input wave packet at (a) $n=0$ and (b) $n=1$. The rest of the parameters are as in Fig.~3. White areas correspond to the parameter region for which the transfer probability into the ring is higher than 97$\%$. Solid curves represent the corresponding threshold velocities shown in Fig.~2.}
\label{fig4}
\end{figure}

\subsection{Extraction protocol}

Let us now turn to the discussion of the extraction process, as depicted in  Fig.~1(a) (center). 
The atoms can be extracted from the ring by applying the same protocol as for injection but simply exchanging the role of the curved and ring waveguides. 
In fact, the extraction protocol corresponds to the time reversal sequence of the injection protocol. Thus, and due to the unitary evolution of the Schr\"odinger equation, all the results that we have obtained for the injection protocol also apply for the extraction protocol.

Taking into consideration that (i) the ring perimeter is much larger than the longitudinal width of the atomic wave packets (so the atomic matter wave propagating along the ring is well localized in a small segment of the ring) and that (ii) the waveguides coupled to the ring/s can be switched on or off at will by simply turning on or off the laser field that generates them, e.g., as it is the case when the dipole waveguide is generated by means of a cylindrical microlens, the output waveguide of the extraction protocol could be part of a secondary ring to be used as an accumulator or storage ring for cold atoms, being repeatedly supplied by the first ring.
In addition, it could be possible to choose the parameters for the ring geometry and the injection/extraction protocol in such a way that for an initially broad atomic velocity distribution in the first ring, only atoms with low enough longitudinal velocity, such that they fulfill the adiabaticity condition, were transfered to the secondary ring. Rethermalization during a round trip in the first ring would deliver additional low-velocity atoms into the secondary ring, presenting this secondary ring an improved starting condition for further manipulation of the atoms.

\begin{figure}[t]
\resizebox{0.5\textwidth}{!}{
\includegraphics{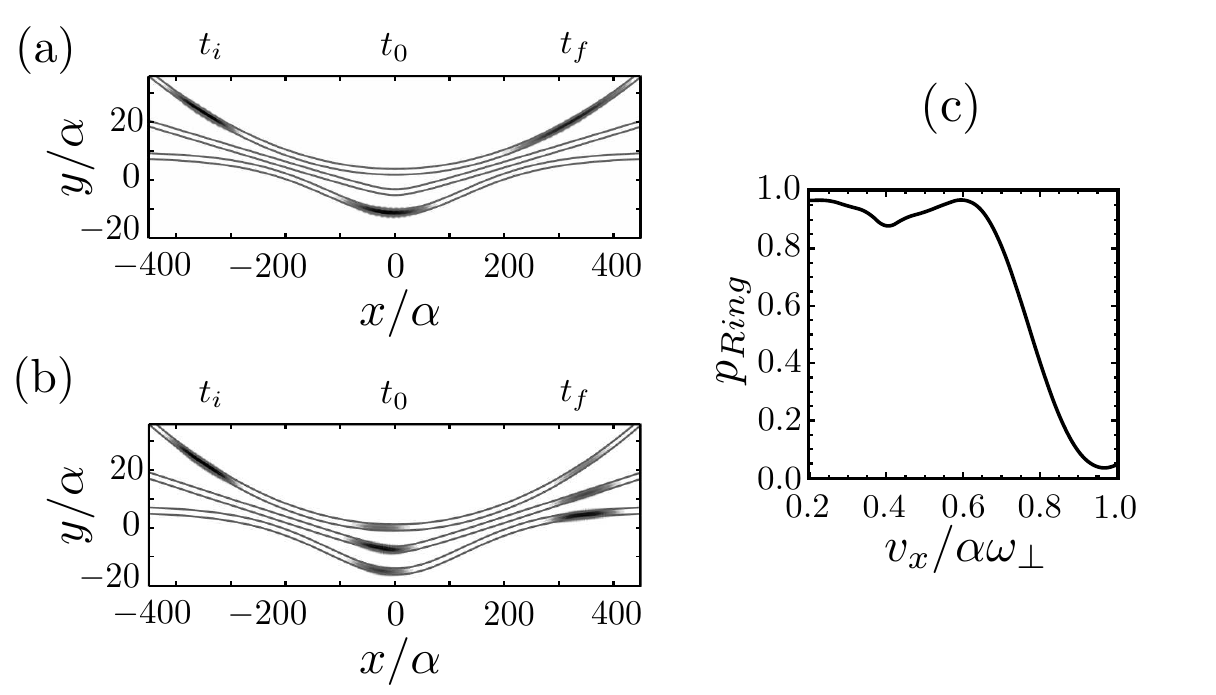}
}
\caption{
Waveguide geometry for a double SAP process including the atomic probability distribution for three different consecutive times with the initial velocity being (a) $v_x=0.3\alpha\omega_\bot$ and (b) $v_x=\alpha\omega_\bot$. 
(c) Final atomic population in the ring as a function of $v_x$. In all cases $n=0$. 
}
\label{fig5}
\end{figure}

\subsection{Velocity filtering protocol}

We turn to the discussion of the velocity filtering configuration shown in Fig.~1(a) (right) and Fig.~5(a), where atoms propagating along the ring ($r=3000\alpha$) are coupled to a particular waveguide geometry designed to perform a double SAP process. Each individual adiabatic passage geometry is characterized by $d_0 =3.9\alpha$ and $x_0=50 \alpha$. At $x=0$ the separation between the waveguides is $7 \alpha$. Figs.~5 (a) and (b) show the evolution of an input atomic wave packet at different times ($t_i$, $t_0$, and $t_f$)  for $v_x=0.3\alpha\omega_\bot$ and $v_x=\alpha\omega_\bot$, respectively, and $n = 0$ in both cases. 
In (a), the double SAP is performed with high efficiency and the atomic wave packet continues to propagate along the ring after the filtering section, whereas in (b)
the input velocity is too high resulting in the spreading of the atomic wave packet among the waveguides. 
Fig.~5(c) shows the final population in the ring as a function of the initial longitudinal velocity showing that slow atoms are able to adiabatically follow the spatial dark state and return to the ring trap with high fidelity, while faster atoms that do not fulfill the adiabaticity condition spread among the three waveguides, thus reducing the final population in the ring. 
One can estimate the expected values of the atomic velocities that will be filtered by means of this geometry for realistic ring geometries: assuming $^{87}$Rb atoms in an optical ring potential with radius of $\sim 1~$mm and trapping frequency $\omega_{\bot} \sim 2\pi \times 1\,$kHz, we obtain $\alpha\omega_\bot \sim 2\,$mm$/$s. Therefore, based on Fig.~5(b), it should be possible to selectively remove atoms with velocities higher than $v_x \sim 0.6\alpha\omega_\bot = 1.2\,$mm/s.
  
  
\section{Conclusions}
  
To conclude, we have presented a coherent technique for the injection of neutral atoms into, extracting them out of, and velocity filtering them inside a ring dipole trap.
The technique is based on adiabatically following the transverse spatial dark state of the system. Explicit conditions for the spatial adiabatic passage between the waveguides and the ring as a function of the initial longitudinal atomic velocity and on the initial population distribution among the transverse vibrational states have been discussed. The performance of our proposal has been checked by numerical integration of the corresponding 2D Schr\"odinger equation. 

Worth to note, this article has been focused on controlling the transfer of a single-atom between the outermost waveguides of a system composed of two dipole waveguides and a ring trap by means of the SAP technique. However, one could also assume a thermal cloud of atoms as the input beam such that only those atoms of the cloud whose longitudinal velocity is lower than the actual threshold velocity would propagate into the ring. Subsequent thermalization into the ring would give rise to a cooler atomic distribution. In addition, it would be also very interesting to discuss the velocity filtering protocol for an input beam consisting of a BEC or a matter-wave soliton and investigate the role of the non-linearity, e.g., whether it could prevent transverse SAP.

\section*{Acknowledgments}

The authors gratefully acknowledge financial support through the Spanish MICINN contract FIS2011-23719, the Catalan Government contract SGR2009-00347, and from the German Academic Exchange Service (DAAD) (Contract No. 0804149). RME acknowledges financial support from AP2008-01276 (MECD).

\end{document}